\documentclass[journal]{IEEEtran}
\IEEEoverridecommandlockouts
\usepackage{cite}
\usepackage{amsmath,amssymb,amsfonts}
\usepackage{algorithmic}
\usepackage{graphicx}
\usepackage{textcomp}
\usepackage{xcolor}
\usepackage{array,multirow}
\usepackage{rotating}
\usepackage{ifpdf}
\usepackage[utf8]{inputenc}
\usepackage[T1]{fontenc}
\usepackage{colortbl}
\usepackage{array,multirow}
\usepackage{tabularx,booktabs,lipsum,graphicx}
\usepackage{rotating}
\usepackage{enumitem}
\usepackage{fixltx2e}
\usepackage{cleveref}
\usepackage{arydshln}
\usepackage{float}
\usepackage[outdir=./]{epstopdf}
\def\BibTeX{{\rm B\kern-.05em{\sc i\kern-.025em b}\kern-.08em
    T\kern-.1667em\lower.7ex\hbox{E}\kern-.125emX}}
\begin{document}
	\bstctlcite{IEEEexample:BSTcontrol}
	\bibliographystyle{IEEEtran}
\title{Comparison of security margin estimation methods under various load configurations}


\author{Hannes~Hagmar,~\IEEEmembership{Student Member,~IEEE,}
        Robert~Eriksson,~\IEEEmembership{Senior Member,~IEEE,}
        and Le~Anh~Tuan,~\IEEEmembership{Member,~IEEE}

\thanks{
The work presented in this paper has been financially supported by Energimyndigheten (Swedish Energy Agency) and Svenska kraftnät (Swedish National Grid) within the SamspEL program with project number 44358-1.}
}

\markboth{}%
\MakeLowercase

\maketitle

\begin{abstract}

The post-contingency loadability limit (PCLL) and the secure operating limit (SOL) are the two main approaches used in computing the security margins of an electric power system. While the SOL is significantly more computationally demanding than the PCLL, it can account for the dynamic response after a disturbance and generally provides a better measure of the security margin. In this study, the difference between these two methods is compared and analyzed for a range of different contingency and load model scenarios. A methodology to allow a fair comparison between the two security margins is developed and tested on a modified version of the Nordic32 test system. The study shows that the SOL can differ significantly from the PCLL, especially when the system has a high penetration of loads with constant power characteristics, or a large share of induction motor loads with fast load restoration. The difference between the methods is also tested for different contingencies, where longer fault clearing times are shown to significantly increase the difference between the two margins. 

\end{abstract}
\begin{IEEEkeywords}
Dynamic security margins, load modelling, security assessment, security margin estimation
\end{IEEEkeywords}

\section{Introduction}
\label{sec:Introduction}

Electric power systems are generally operated according to the N-1 contingency criterion, meaning that the system should be able to withstand the loss of any single system component, such as a transmission line or a generating unit, without losing stability. A system capable of handling this is said to be \textit{secure} \cite{DefinitionVS}. However, even when the system is secure for a given operation condition, system operators are also required to know how far the system can move from its current operating point and still remain secure. To estimate this, system operators continuously compute \textit{security margins}, which in turn represents the available transmission capacity in the system. 

There exist two main approaches to compute the security margins of a power system: the post-contingency loadability limit (PCLL) and the secure operating limit (SOL) \cite{Cutsem1998}. The PCLL is evaluated by estimating the loadability limit of a post-contingency operating point, where a solution path is traced by iteratively increasing the system stress until the system's critical point is reached. The characteristics of the iteratively increased system-stress in the post-contingency setting are similar to that of the slow load restoration that typically follows in a long-term voltage stability event. Thus, it is common that the PCLL is also simply referred to as the voltage security margin (VSM). 

An alternative measure of the security margin is the SOL, which refers to the most stressed, \textit{pre-contingency} operating state in which the system can withstand a specified set of contingencies. The SOL can account for the dynamic response after a contingency and provides a generally more accurate measure of the actual security margin of the system. However, the SOL has been comparatively less documented in the literature, likely due to the practical difficulties required in its estimation. The SOL requires numerous time-domain or quasi steady-state (QSS) simulations to trace the security limit for a set of different contingencies, a task that is generally too time-consuming to meet the near real-time monitoring requirements of system operators. In applications where the SOL is mainly used in monitoring voltage stability, it is also referred to as the dynamic voltage security margin (DVSM). 

Several studies have attempted to provide solutions to the high computational effort required in estimating the SOL. Methods using QSS simulations, as found in \cite{Cutsem1999}, can reduce the simulation time significantly compared to full time-domain simulations, but cannot fully incorporate the impact of short-term and transient effects. In \cite{CUTSEM2005b}, a method that combined QSS and time-domain simulations was proposed to include the impact of short-term effects. In \cite{Sittithumwat2002,AMJADY2003,VAKILBAGHMISHEH2007}, various machine learning approaches based on training neural networks were proposed to allow real-time estimation of the DVSM. In \cite{Hagmar2020IET}, a combined methodology based on validating the estimations of neural networks with actual time-domain simulations was proposed to overcome the robustness issues that are commonly related to machine learning methods.

Despite ongoing efforts in improving the computational efficiency of the SOL, the circumstances when the SOL is to prefer to the PCLL have been relatively unexplored in the literature. In \cite{Dobson1994}, it was shown that if a system starts at a stable equilibrium and is slowly stressed towards a critical point without encountering oscillations or other limit-induced events (e.g. reactive power limits for generators), static estimation methods are sufficient to locate the exact critical point experienced by the dynamic system. Thus, in such circumstances, the security margin computed by PCLL would essentially be equal to the one computed using the method of the SOL. In \cite{Hagmar2020IET}, a theoretical comparison between the DVSM and VSM was performed, where the difference between the two measures was illustrated using a concept called "transient $P\textrm{-}V$ curves". The study highlighted the importance of load restoration dynamics on the difference between the two methods but provided no numerical results. In \cite{Chowdhury2000}, the SOL was numerically compared to another type of security margin computed by static $V$-$Q$ curves, in which variations in the reactive power injection at a bus would affect the voltage at that same bus. The study concluded that the SOL (or generally a dynamic simulation approach) is a superior method compared to $V$-$Q$ curves, but since the two methods are so conceptually different, the results of the two methods could not directly be compared. In \cite{Cutsem1999}, a SOL computed by QSS simulations was compared to the PCLL, where primarily the impact of post-disturbance control was investigated. However, as will later be discussed, QSS simulations cannot fully incorporate the short-term effects after a disturbance and no variations in the load composition was analyzed in the study.

To address the above-mentioned lack of an accurate comparison between the methods, this study aims to provide a comprehensive analysis of the difference between the PCLL and the SOL. The purpose is to isolate the root cause of the difference between the two security margin methods, referring to that the SOL could better account for the system's dynamic response after a disturbance, and to examine under what circumstances this causes the results of the methods to differ. The study is, to the authors' knowledge, the first in developing a methodology which allows the two margins to be fairly compared for a large range of different load configurations and disturbance scenarios. 

The main contributions of this study are the following: 

\begin{itemize}
    \item A methodology to allow a fair comparison between the PCLL and the SOL is developed. The purpose of the developed methodology is to ensure that the difference between the computed margins are due to \textit{actual} differences of the security margin approaches, rather than differences in how the simulations are conducted. 
    \item An extensive numerical comparison between the SOL and the PCLL under a range of both static and dynamic load model configurations is performed. Different fault scenarios are examined and discussed in the study. The purpose is to examine under what circumstances that the SOL are preferable to the PCLL. 
\end{itemize}

The rest of the paper is organized as follows. In Section II, the security margin definitions for the SOL and the PCLL are presented. In Section III, the methodology used in computing the margins is presented along with the simulation platform and the adaptations used to allow a fair comparison between the security margins. Results and discussion are presented in Section IV. Concluding remarks are presented in Section V. 

\section{Security margin definitions}
In this section, a theoretical comparison between PCLL and the SOL is presented. The conceptual difference of computing the two security margins is first presented, which is followed by a theoretical analysis 
of the circumstances and the instability mechanisms that can cause the two methods to differ. 

\subsection{PCLL versus SOL}

\begin{figure}
	\begin{center}\vspace{-0.2cm}
		\includegraphics[width=7.8cm]{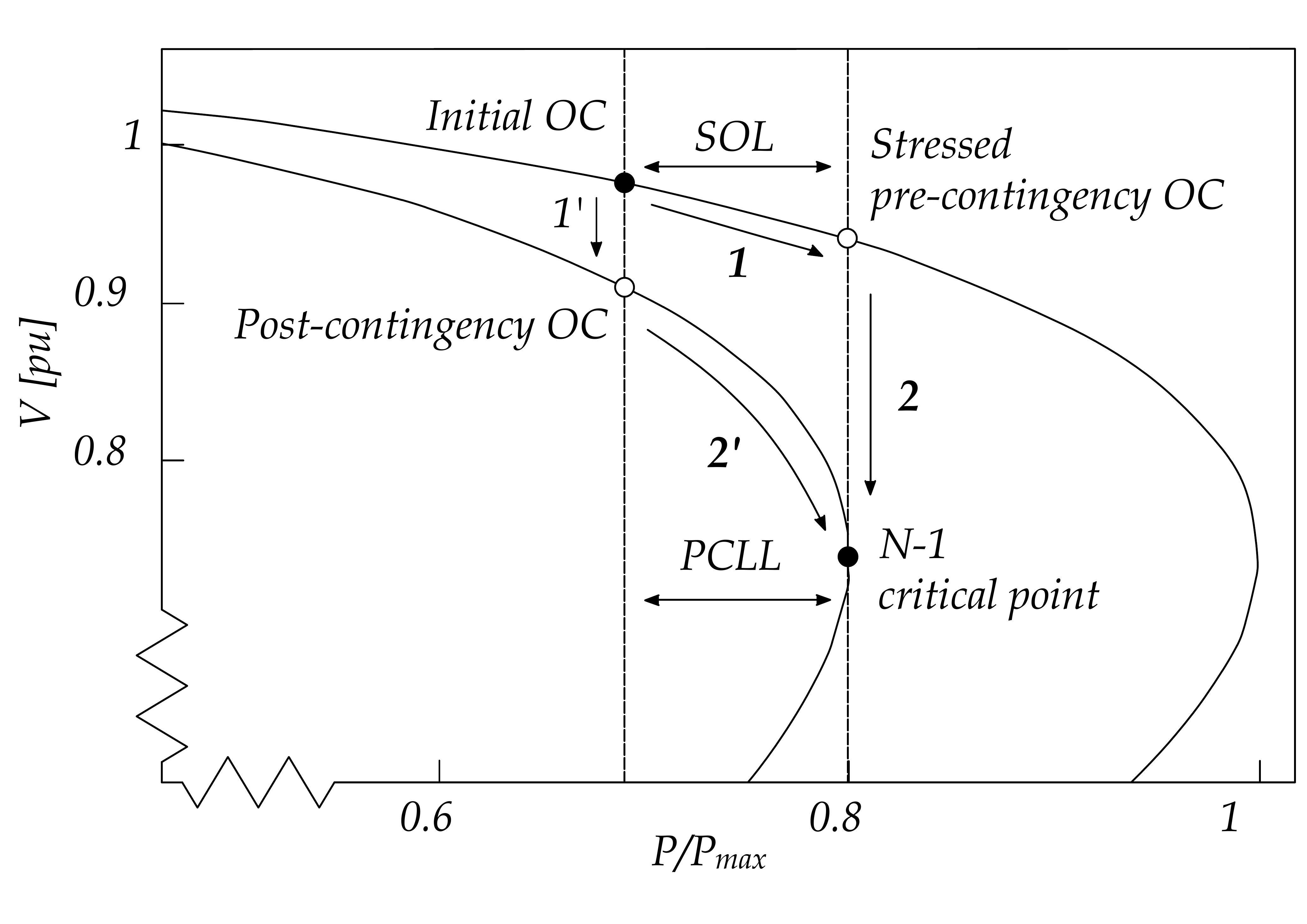}    
		\vspace{-0.6cm}
	\caption{Security margin estimation for PCLL and SOL. $\textbf{1'}$ and $\textbf{2'}$ illustrates the computation path for the PCLL; $\textbf{1}$ and $\textbf{2}$ illustrates the computation path for the SOL.} 
		\label{fig:PCLL_vs_SOL}
		\vspace{-0.8cm}
	\end{center}
\end{figure}

The security margin estimation processes for the PCLL and the SOL are illustrated in Fig. \ref{fig:PCLL_vs_SOL}. Pre-contingency and post-contingency $P\textrm{-}V$ curves are used where a receiving end voltage in a stressed area is a function of an increasing (active) power transfer from the system to this receiving end. An initial, unstressed, operating condition (OC) is the starting point for the security margin estimation. The security margin is then defined as the change in loading from the initial OC to the $N$-$1$ critical point. It should be noted that in actual applications, the limit is often smaller due to the other stopping criteria such as too low system voltages. However, for better illustration purposes, the former limit is used here. 

In PCLL estimation, a post-contingency operating point is found by first introducing a contingency on the initial OC, which is typically followed by solving the resulting power flow study. This is illustrated in Fig.~\ref{fig:PCLL_vs_SOL} by moving along arrow $\boldsymbol{1'}$. The security margin is then traced along the solution path by iteratively and step-wise increasing the system stress until the critical point is reached, moving along the arrow $\boldsymbol{2'}$. Parameterized continuation methods based on static load flow solutions, generally referred to as continuation power flow (CPF), are commonly used to avoid convergence problems close to the critical point of the system \cite{Ajjarapu1992}. The distance between the pre-contingency operating point and the $N$-$1$ critical point represents the PCLL. 

In the estimation of the SOL, the dynamic security of the system is tested with an increasing stress level, illustrated by arrow $\boldsymbol{1}$ in Fig.~\ref{fig:PCLL_vs_SOL} \cite{Cutsem1999}. For every new stressed pre-contingency OC, the system response following the disturbance is studied. The final, pre-contingency OC that is tested and can provide a stable operating point is illustrated by moving along arrow $\boldsymbol{2}$ in Fig.~\ref{fig:PCLL_vs_SOL}. The state transition following arrow $\boldsymbol{2}$ can generally not be computed using static methods as it can result in numerical divergence. Instead, methods based on dynamic (or QSS) simulations are generally required. The increased loadability between the initial OC and the most stressed pre-contingency OC that can still handle a dimensioning contingency without causing instability represents the SOL. 

\subsection{System dynamics and instability mechanisms}
\label{sec:System_dynamics}
Loads are often recognized to maintain a constant power characteristics in a long-term \textit{system} perspective, but do not generally behave as such following a disturbance. Assuming a sudden voltage change, loads will initially drop according to their instantaneous characteristics \cite{Pal1992}. Then, the impedance or the drawn current is adjusted to restore the load to its original level; a process that can be exemplified by the automatic changes in the slip of induction motors or by changes in tap positions to increase the voltage for loads behind load tap changers (LTCs). The overall load restoration after a disturbance is generally assumed to act significantly slower than the dynamics of other system components, such as the dynamics of generators and excitation systems. The PCLL is based on this time-scale decomposition, where short-term dynamics, such as the slip of induction motors, or generator and excitation dynamics, are assumed to be in equilibrium. Using this assumption, the loadability limit of the post-disturbance system can be found even though only static estimation methods are used to trace the security margin. 

\begin{figure}
	\begin{center}\vspace{-0.3cm}
		\includegraphics[width=7.8cm]{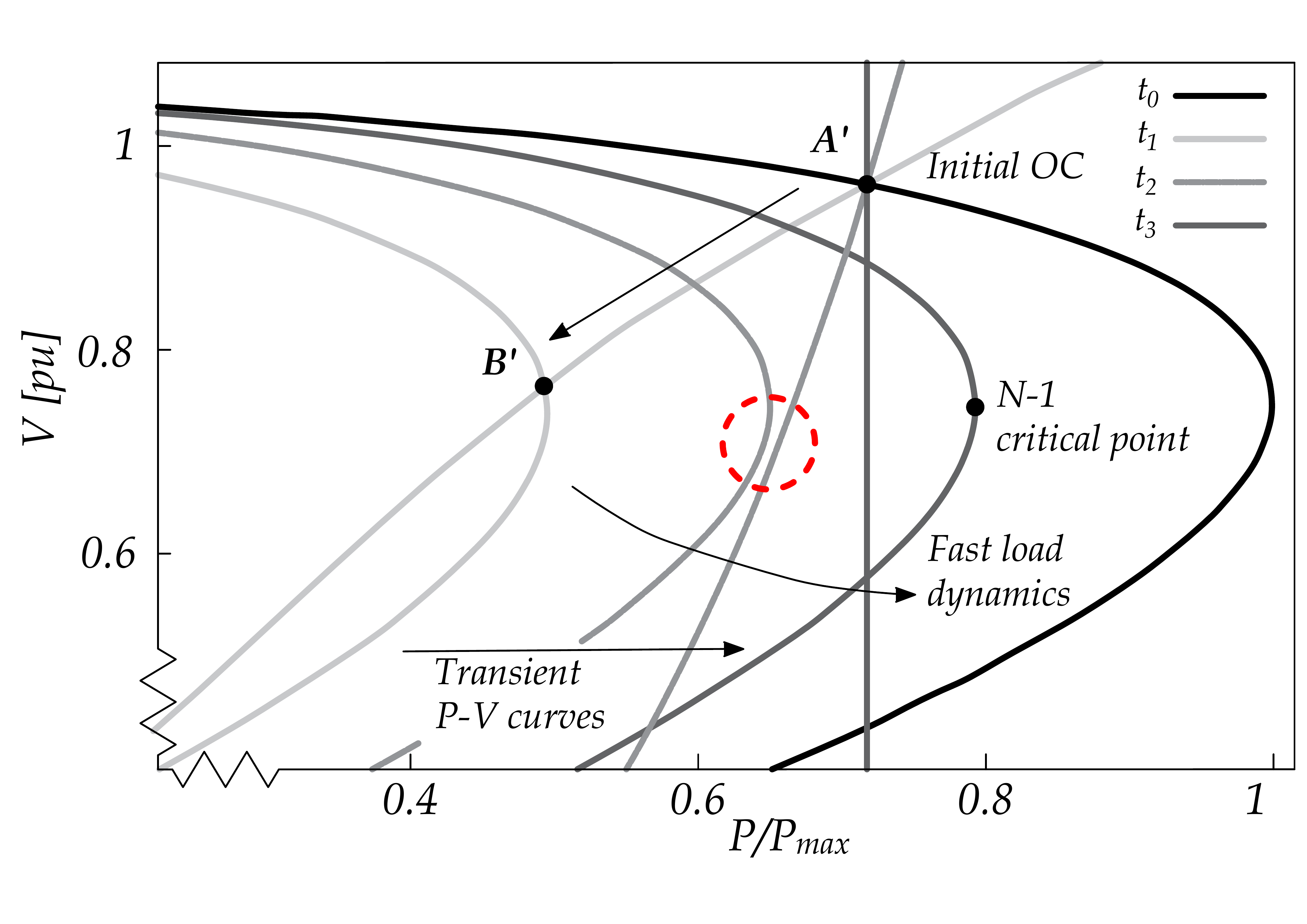}    
		\vspace{-0.7cm}
		\caption{Transient $P\textrm{-}V$ curves illustrating a short-term instability event \cite{Hagmar2020IET}}
		\label{fig:Transient_PVcurvesB}
		\vspace{-0.8cm}
	\end{center}
\end{figure}

In \cite{Pal1992} and \cite{Hagmar2020IET}, the concept of \textit{transient} $P\textrm{-}V$ curves was used to allow visualization and analysis of short-term dynamics using $P\textrm{-}V$ curves. In the analysis, the post-disturbance $P$-$V$ curve is not fixed in time, but is allowed to be affected by short-term system dynamics of, for instance, excitation systems. Nor is the load curve fixed in time, which allows the load restoration that follows after a disturbance to be illustrated. In Fig. \ref{fig:Transient_PVcurvesB}, transient $P\textrm{-}V$ curves are used to illustrate a system that, when assuming the short-term dynamics are in equilibrium, could appear to be secure. However, when the short-term dynamics are taken into account, there is a loss of post-disturbance equilibrium of the short-term dynamics, and the disturbance would in fact cause the system to become unstable. The load restoration curves and the transient $P\textrm{-}V$ curves are illustrated using different shades of grey, where a lighter shade indicates closer in time after the disturbance. The time just after a disturbance is indicated by $t_1$; $t_2$ relates to a short time after the disturbance; $t_3$ relates to the time when all short-term dynamics are in equilibrium. The load is assumed to have long-term constant power characteristic, but just after a disturbance, the load will initially change to a constant impedance characteristic. Then, by fast load restoration, the load is quickly restored to the pre-disturbance level.

The initial OC is found in point $\boldsymbol{A'}$. Just after the disturbance, the short-term dynamics of system components such as generators or excitation systems will not yet have stabilized, which has the effect of shifting the post-disturbance $P$-$V$ curve to the left. As a result of the initial load characteristics and the shifted transient $P$-$V$ curve, the operating point moves along the arrow to operating point $\boldsymbol{B'}$. After this transition, both the load and the post-disturbance $P$-$V$ curve are shifted towards their stable counterparts. However, due to the fast load dynamics in this case, there exists no intersection between the load curve and the post-disturbance $P$-$V$ curve at $t_2$ (the area indicated with the red dotted circle), and without any emergency control actions, the system stability would be lost. Thus, despite the fact that the post-disturbance $P$-$V$ curve and the load characteristic at $t_3$ still intersect in this case, the system would have become unstable. 

Instability caused by the short-term dynamics that follows a disturbance can generally be divided into three different instability mechanisms \cite{Cutsem1998, DefinitionVS}: 
\begin{itemize}
    \item Loss of post-disturbance equilibrium of short-term dynamics: Typically exemplified by the stalling of induction motors after a disturbance causing the transmission impedance to increase. Due to the increased transmission impedance, the mechanical and electrical torque curves of the motor may not intersect, causing the system to lack a post-disturbance equilibrium, similar to the case illustrated in Fig. \ref{fig:Transient_PVcurvesB}. 
    \item Lack of attraction towards the stable post-disturbance equilibrium of
    short-term dynamics: Typically exemplified by transient angle instability and the loss of synchronism by one (or several) generators following a too slow fault clearing. 
    \item Oscillatory instability of the post-disturbance equilibrium: Typically exemplified by rotor angle stability, in which the equilibrium between electromagnetic torque and mechanical torque of synchronous machines in
    the system is disturbed. Instability may be caused by increasing angular swings of some generators leading to their loss of synchronism with other generators \cite{DefinitionVS}.
\end{itemize}

Typically, time-domain simulations are required to capture the short-term dynamics after a large disturbance. SOLs computed using QSS simulations can typically not account for the short-term dynamics that follows after a disturbance, and are thus better suited to only monitor long-term voltage instability phenomena. However, combinations of time-domain simulations and QSS, as proposed in \cite{CUTSEM2005b}, can use time-domain simulations to model the system during the short-term period following a disturbance, followed by QSS simulations used to simulate the long-term interval after the short-term effects are finalized. However, short-term instability may also be induced by long-term dynamics, where the system degradation caused by long-term instability eventually can trigger the above mentioned short-term events \cite{Cutsem1998}. It should be noted that SOLs computed by combinations of time-domain simulations and QSS, as proposed in \cite{CUTSEM2005b}, cannot capture this type of events. 

\section{Methodology for security margin computations}

In this section, the methodology used in the comparison between the PCLL and the SOL is presented. The load models and a description of the test system are presented along with the required adaptions. Finally, the methodology used to allow a fair comparison of the PCLL and the SOL is presented. 

\subsection{Load models}
The power consumption of loads are generally affected by the system voltages and different load models are often used to characterize this relationship. A traditional load model used in both static and dynamic stability analysis is the ZIP model, which is made up of three components: constant impedance ($Z$), constant current ($I$), and constant power ($P$). The characteristics of the ZIP model is given by \cite{Cutsem1998}: 
\begin{subequations}
    \begin{align}
        P = zP_0\left[a_P\left(\frac{V}{V_0}\right)^2+b_P\frac{V}{V_0}+c_P\right] \\
        Q = zQ_0\left[a_Q\left(\frac{V}{V_0}\right)^2+b_Q\frac{V}{V_0}+c_Q\right]
    \end{align}
\end{subequations}
where $a_P+b_P+c_P$ = $a_Q+b_P+c_Q$ = $1$, $P_0$ and $Q_0$ are the real and reactive powers consumed at a reference voltage $V_0$, given that $z$ = $1$. $V$ is the actual voltage and $z$ is a variable indicating the actual loading of the system \cite{Cutsem1998}. The constants $a_x$, $b_x$, and $c_x$ represents the share of constant impedance, constant current, and constant power of the load, respectively. 

\begin{figure}
	\begin{center}\vspace{-0.9cm}
		\includegraphics[width=7.0cm]{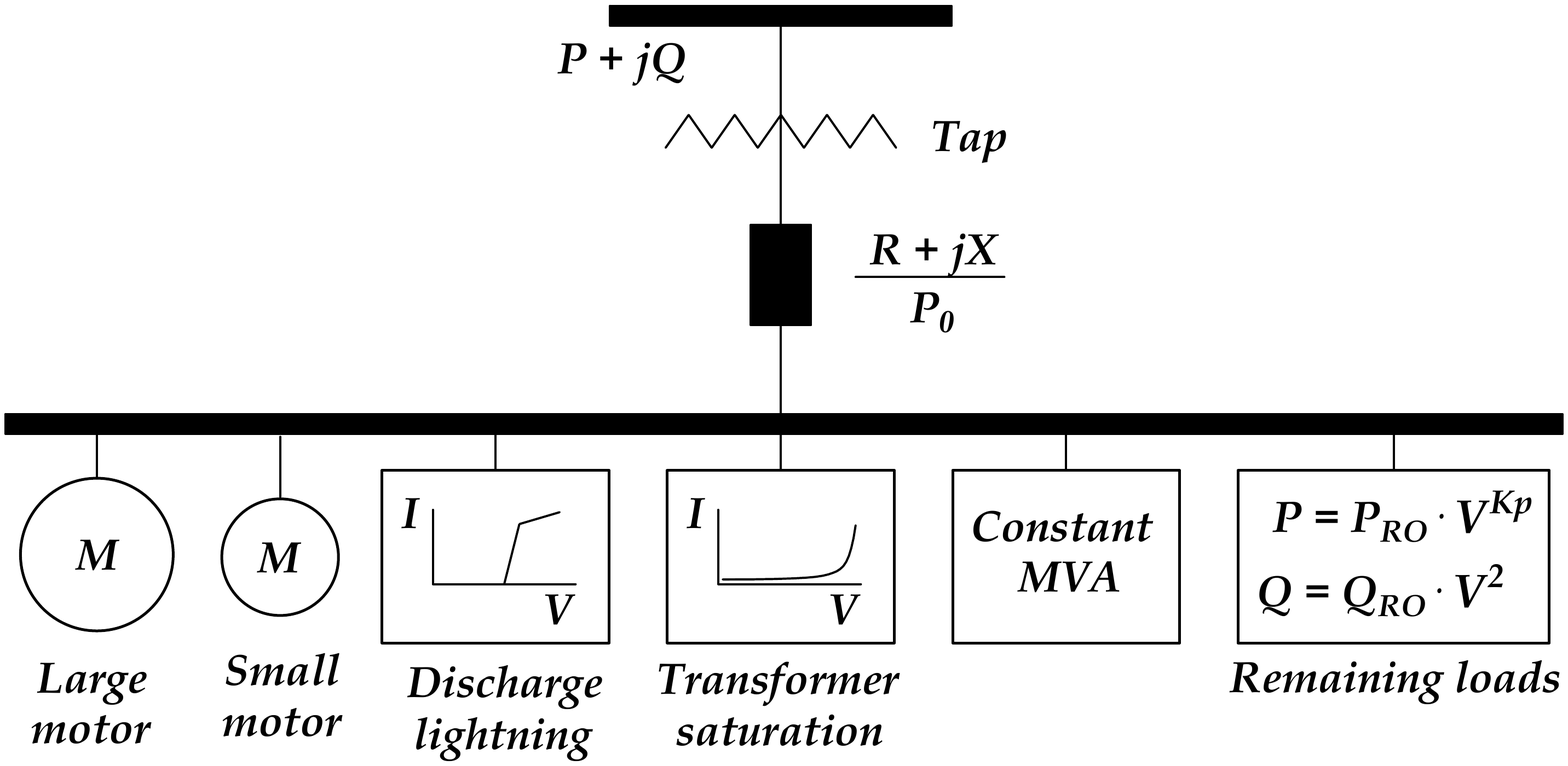}    
		\vspace{-1.1cm}
		\caption{Overview of the complex load model (CLOD) \cite{PAGV2}.}
		\label{fig:CLOD_overview}
		\vspace{-0.8cm}
	\end{center}
\end{figure}

Although simple and widely used in security analysis \cite{Arif2018}, the ZIP model cannot model any dynamic behaviour of the loads themselves. The significance of induction motor loads and other fast-acting dynamic loads are often highlighted in system stability studies. Induction motors (IM) are characterized by fast load restoration dynamics (often in the time frame of a second) and have a high reactive power demand. Induction motors are also prone to stalling, which may cause the motor to draw high reactive currents from the grid, resulting in a deteriorating effect on the system stability \cite{Cutsem1998}. In PSS\textregistered{}E, a complex load model (CLOD) can be used to represent a bundled mix of loads with different dynamic load characteristics into a single model. In Fig.~\ref{fig:CLOD_overview}, a schematic of the CLOD model is presented. The transformer and feeders connecting the system bus to the load bus are modelled as a single impedance. At the load bus, different percentages of large and small IMs, discharge lighting loads, transformer saturation, and constant power loads can be modelled. The remaining part of the load is modelled as a polynomial load where the voltage dependency of the active load is controlled through a constant $K_p$. The performance curves of the two motor models, the discharge lightning model, and the transformer saturation model is further detailed in~\cite{PAGV2}.

\subsection{Simulation test system}

\begin{figure}[t!]
	\begin{center}\vspace{-0.6cm}
		\includegraphics[width=5.8cm]{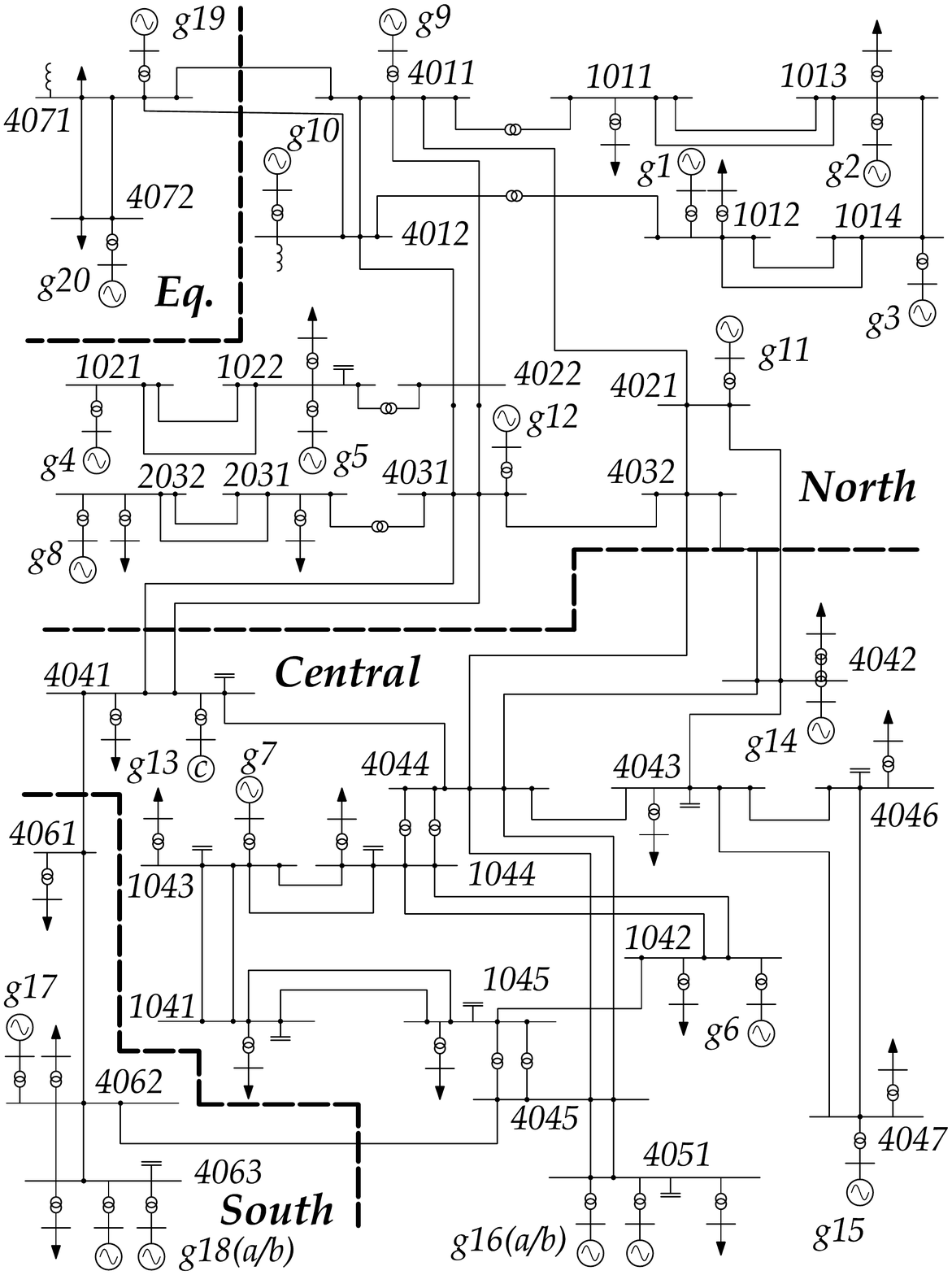}    
		\vspace{-0.5cm}
		\caption{One-line diagram of the modified Nordic32 system \cite{CutsemTestSystem2020}} 
		\label{fig:Nordic32}
		\vspace{-0.5cm}
	\end{center}
\end{figure}

All simulations have been tested on the slightly modified version of the Nordic32 test system, detailed in \cite{CutsemTestSystem2020}. The main characteristic of the system is sensitivity towards long-term voltage instability, although the system can exhibit other types of instabilities as well. A one-line diagram of the test system is presented in Fig.~\ref{fig:Nordic32}. The security margins are computed by increasing the loading in the area "\textit{Central}", while the generation in the area "\textit{North}" is increased by a corresponding quantity. The starting point for all scenarios is the secure "operating point B" as defined in~\cite{CutsemTestSystem2020}. 

\begin{figure}
	\centering
	\vspace{-1.8cm}
	\includegraphics[width=7.2cm]{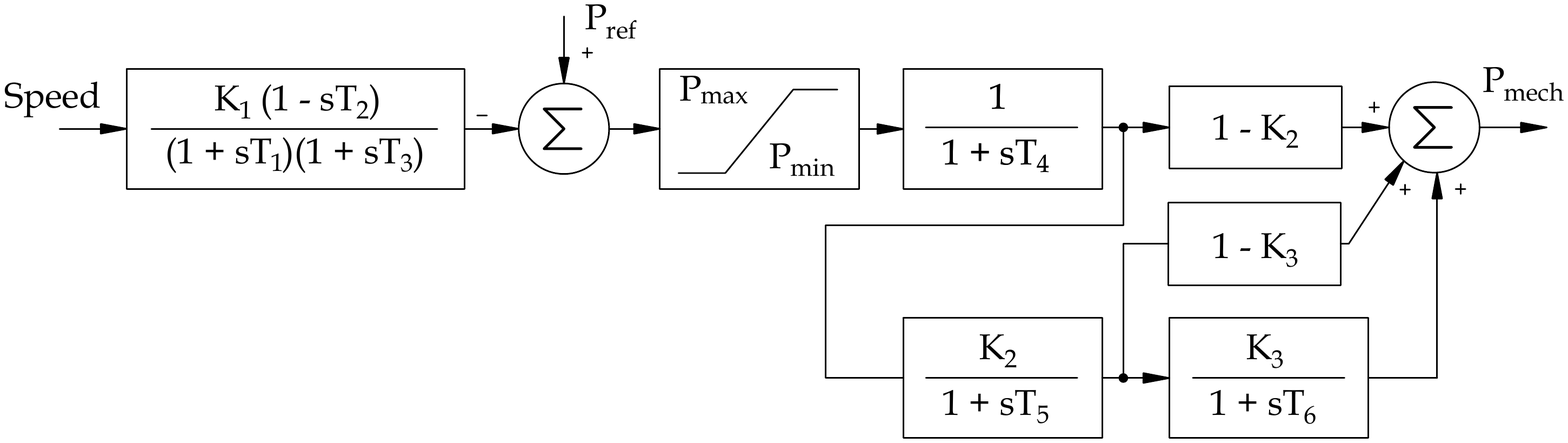}    
	\vspace{-1.7cm}
	\caption{Block diagram of the IEESGO Model \cite{PAGV2}}
	\label{fig:IEEEGO}
	\vspace{-0.6cm}
\end{figure}

All simulations were carried out using PSS\textregistered{}E version 35.0. To ensure numerical stability during the dynamical simulation runs, a short integration step of 0.0005 seconds was used in the simulations. In certain sensitive scenarios, such as when the simulations resulted in a non-converging dynamic simulation, the integration step was at times varied to provide a converging simulation. To more easily allow direct adjustments of the generator power during dynamical simulations, the governor models were changed to the more generic governor model of IEESGO \cite{PAGV2}, illustrated in Fig.~\ref{fig:IEEEGO}. To allow almost instantaneous changes in the mechanical input to generators given the governor reference changes, all controller lags, delays, and constant parameters was chosen zero, except for the governor lag constant $T_3$ which was given a value of~0.01. 

\subsection{Methodology and adaptations}
To compare the two conceptually different methods of PCLL and SOL is not trivial, as one is computed using a static model of the system, while the other is generally estimated using a dynamic model. To ensure that the difference in the computed security margins was not caused by differences in how the simulations were conducted, but rather by the fact that the SOL could better account for the system's dynamic response after a disturbance, a few adaptations of the methods were required. Instead of using CPF methods to compute the PCLL, we adopted a method that \textit{slowly} ramps up the system stress in a dynamic simulation setting; an approach similar to the one used in \cite{CutsemTestSystem2020}. This approach allows the PCLL to be performed in a dynamic setting while mimicking how the system stress would have been increased if it would have been performed in a static setting. The advantage by adopting this methodology is that the loading and the generation set points could be increased in the \textit{exact} same way for both the computation of the PCLL and the SOL. 

\subsubsection{Steps for PCLL computation}

The steps used in computing the PCLL were the following: 

    \begin{itemize}
        \item \textbf{Initialization:} The PCLL computation was initialized by applying a chosen contingency in the system in a dynamic simulation from the base case. The dynamic simulation ran until the system was fully stabilized. 
        
        \item \textbf{Increase system stress:} Once the system had stabilized after the initial disturbance, the system stress was increased in small increments of 1 MW, which was distributed among all the loads in the "\textit{Central}" area. To reduce the required simulation time, the system stress was for certain fault scenarios (scenario A and B) initially increased in larger increments (5 MW) up until a total system stress of 300 MW, since such low stress levels were found to not cause instability in the system. The different fault scenarios are further discussed in Section \ref{sec:Contingency_scenarios_and_loading_scenarios}. Simultaneously, the governor reference for all the generation units in the "\textit{North}" area was increased to compensate for the increased load level. The power factor of each load was kept constant. The added load for both the PCLL and SOL were computed as a nominal load increase at 1.0 pu voltage to ensure that the same amount of load was added for both methods regardless of the current load voltage in the dynamic simulation. Increased active line losses caused by the increased system stress were automatically compensated by the generators' governor systems, while the reactive line losses were automatically compensated by the generators' excitation systems.
        \item \textbf{Check stability criterion:} After the increased system stress, the simulation continued to run until the system either stabilized or until a stability criterion was violated. The system was considered unstable if \textit{any} bus voltage in the system was lower than 0.9 pu. Although the modified Nordic32 test system is characterized by sensitivity towards voltage instability, other types of instability can violate the stability criterion. For instance, transient angle instability can cause locally low voltages due to lost synchronism of certain generators. 
        \item \textbf{Re-iterate:} The system stress was increased until the system eventually violates the stability criterion. The difference in loading from the base case to the final stable operating point in the stressed post-contingency system represents the computed PCLL. 
    \end{itemize} 

\subsubsection{Steps for SOL computation}
The SOL was computed similarly to the PCLL, but by instead stressing the system in its pre-contingency configuration and \textit{then} introducing the disturbance. The steps used in the computation of the SOLs were the following: 

    \begin{itemize}
        \item \textbf{Initialize simulation and increase system stress:} When the ZIP-model was used to model the loads, the dynamic simulation was initialized at the start. The system stress was then increased in its \textit{pre-contingency} base case in small increments of 1 MW, in the same way as was done for the PCLL computation in its post-contingency configuration. The small step-size in system stress was chosen to allow the illustration of the security margins using $P\textrm{-}V$ curves. In more general applications, faster methods such as the binary search method \cite{Cutsem1999} or the dual binary search method \cite{Hagmar2020IET}, can otherwise be used to compute the SOL. \textit{Adjustments for CLOD-model:} When using the CLOD-model to model the loads in the system, the increased system stress was required to be added in a static load flow scenario, which was then converted for dynamical studies. This causes a small difference in how the generation in the "\textit{North}" area is increased. For the PCLL and the SOL case for the ZIP-model, the generation is increased by changing the generators mechanical power by incrementally increasing the governor reference. When the generation is increased using the CLOD-model, the active generation is directly changed in the static load flow scenario which creates a small difference of how the system stress is increased. 
        \item \textbf{Introduce disturbance and check stability:} A disturbance was then applied in the system. A final end time of the dynamic simulation of 1000 seconds was used. The system was considered unstable if \textit{any} bus voltage in the system was lower than 0.9 pu at the end of the simulation. The simulation was also stopped in advance if any bus voltage decreased below 0.7 pu (still allowing the system to first stabilize for 20 seconds after the disturbance). 
        \item \textbf{Re-iterate:} If the system was stable, the system was reinitialized to the last pre-contingency state, followed by increasing the system stress by an additional 1 MW and applying the same disturbance. The SOL is then computed from the difference in loading from the base case to the final stable operating point in the stressed post-contingency system.
    \end{itemize} 

\section{Simulation results and discussion}

In this section, the results of the numerical comparison between the PCLL and the SOL is presented. Three different contingency scenarios were tested. The results of PCLL and the SOL computation are presented for each contingency scenario and each load model configuration. 
\subsection{Contingency scenarios and loading scenarios}
\label{sec:Contingency_scenarios_and_loading_scenarios}

\begin{table*}
	\vspace{-0.4cm}
	\centering
	\caption{Computed PCLL and SOL for different loading and contingency scenarios}
	\vspace{-0.25cm}
	\label{table:Results_ZIP}
	\begin{tabular}{cccccccccccccccccccc}
		\toprule
	     && \multicolumn{3}{c}{\textbf{Constant}} &&&&&&&&& \\ 
	     && \textbf{MVA} & \textbf{I} & \textbf{Z} &&&
	     \multicolumn{2}{c}{\textbf{\textit{Scenario A}}} &&& 
	     \multicolumn{2}{c}{\textbf{\textit{Scenario B}}} &&&
	     \multicolumn{2}{c}{\textbf{\textit{Scenario C}}} \\ 
	    \cmidrule(lr){3-5}\cmidrule(lr){8-9}\cmidrule(lr){12-13}\cmidrule(lr){16-17}
        \textbf{Scenario} && (P/Q) & (P/Q) & (P/Q) &&& \textbf{PCLL} & \textbf{SOL} &&& \textbf{PCLL} & \textbf{SOL} &&& \textbf{PCLL} & \textbf{SOL}  \\ 
	    \textbf{number} && [\%] & [\%] & [\%] &&& [MW] & [MW] &&& [MW] & [MW] &&& [MW] & [MW] \\ 
	    \hline
	    \addlinespace
	    1 && 100/0 & 0/0 & 0/100 &&& 372 & 142 &&& 372 & 5 &&& - & - \\ 
	    2 && 95/0 & 5/0 & 0/100 &&& 378 & 245 &&& 378 & 174 &&& - & - \\ 
	    3 && 90/0 & 10/0 & 0/100 &&& 386 & 311 &&& 386 & 282 &&& 150 & - \\ 
	    \addlinespace
	    4 && 80/0 & 20/0 & 0/100 &&& 399 & 398 &&& 399 & 398 &&& 173 & 5 \\ 
	    5 && 50/0 & 50/0 & 0/100 &&& 425 & 417 &&& 425 & 417 &&& 281 & 267 \\ 
	    6 && 95/0 & 5/50 & 0/50 &&& 265 & 152 &&& 265 & 152 &&& - & - \\ 
	    \addlinespace
	    7 && 80/0 & 20/50 & 0/50 &&& 393 & 363 &&& 393 & 255 &&& 126 & 9 \\ 
	    8 && 50/0 & 50/50 & 0/50 &&& 418 & 411 &&& 418 & 411 &&& 259 & 259 \\ 
	    9 && 0/0 & 100/0 & 0/100 &&& 450 & 441 &&& 450 & 440 &&& 336 & 268 \\ 
	    \addlinespace
	    10 && 0/0 & 50/0 & 50/100 &&& 473 & 463 &&& 473 & 440 &&& 375 & 290 \\ 
	    11 && 0/0 & 20/0 & 80/100 &&& 492 & 480 &&& 492 & 475 &&& 393 & 301  \\ 
	    12 && 0/0 & 0/0 & 100/100 &&& 497 & 493 &&& 497 & 489 &&& 408 & 305 \\ 
		\bottomrule
		\vspace{-0.9cm}
	\end{tabular}
\end{table*}

The following contingency scenarios were examined: 
\begin{itemize}
    \item \textbf{Scenario A:} A three-phased fault for 40 milliseconds, followed by tripping the faulted line. The faulted line is the one connecting the two areas "\textit{North}" and "\textit{Central}" between bus 4032 to bus 4044. 
    \item \textbf{Scenario B:} A longer three-phased fault for 100 milliseconds, followed by tripping the faulted line. The faulted line is the one connecting the two areas "\textit{North}" and "\textit{Central}" between bus 4032 to bus 4044. 
    \item \textbf{Scenario C:} Tripping the of generator "\textit{g16b}" located at bus 4051 in the "\textit{Central}" area. 
\end{itemize}

For each of the contingency scenarios, different combinations of the ZIP-load were tested for both the PCLL and the SOL. In addition, the SOL was computed for different compositions when the CLOD model was used to model the loads in the system. The CLOD model was found to be numerically unstable for longer fault clearing times. Thus, we only provide a comparison of the results for Scenario A with a fault clearing time of 40 milliseconds.

\subsection{Simulation results}

\begin{figure}
	\begin{center}\vspace{-2.8cm}
		\includegraphics[width=7.2cm]{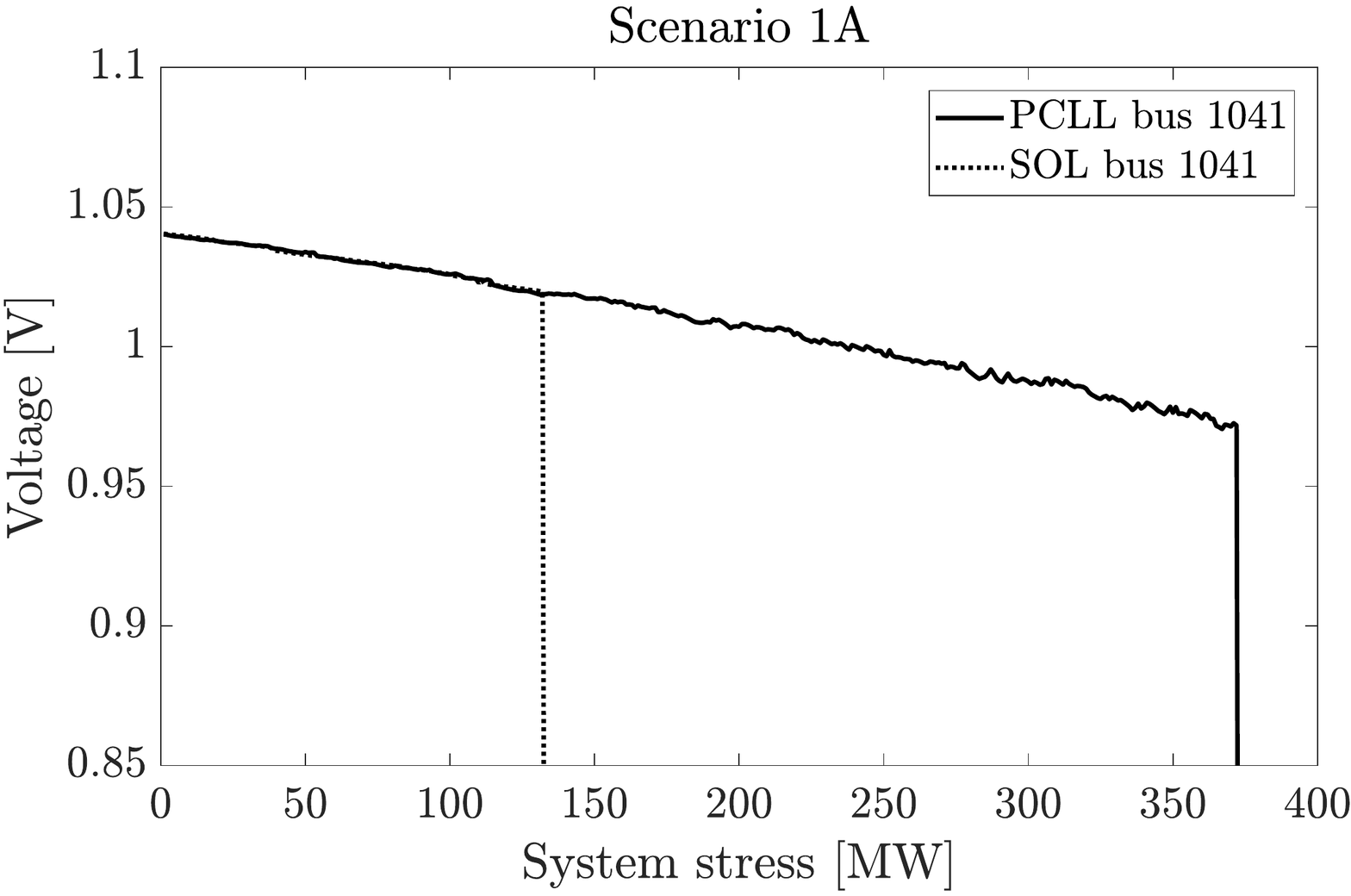}    
		\vspace{-3.15cm}
		\caption{$P\textrm{-}V$ curves computed at bus 1041 for scenario 1A}
		\label{fig:1041_Scenario1A}
		\vspace{-0.90cm}
	\end{center}
\end{figure}

The PCLL and SOL results for each scenario and each load configuration using the ZIP-model are presented in Table~\ref{table:Results_ZIP}. The SOL results for scenario A and different configurations of the CLOD model are presented in Table~\ref{table:Results_CLOD}. The largest difference between the PCLL and SOL is found for cases with a high share of constant power characteristic of the active part of the loads. For instance, for scenario 1A with fully constant power characteristics for the active part of the load and fully constant impedance characteristics for the reactive part of the load, the SOL was only 142 MW, while the PCLL was found to be 372 MW. The difference between the two security margin methods then reduces rapidly with a decreasing level of constant power characteristics on the active part of the load. Already at slightly lower levels of constant power loads, for instance, in Scenario 4A, the difference between the SOL and the PCLL becomes close to negligible. In Fig.~\ref{fig:1041_Scenario1A}, the post-disturbance $P\textrm{-}V$ curves of the transmission side of bus 1041 are illustrated, respectively, for Scenario 1A. The $P\textrm{-}V$ curves are computed by sampling the voltage magnitude when the system had stabilized after each dynamic simulation. Here, with a fully constant power characteristic of the active part of the loads, the $P\textrm{-}V$ curves are almost identical for both the PCLL and the SOL up until the collapse point for the SOL.

\begin{figure}
	\begin{center}\vspace{-3.1cm}
	\includegraphics[width=7.2cm]{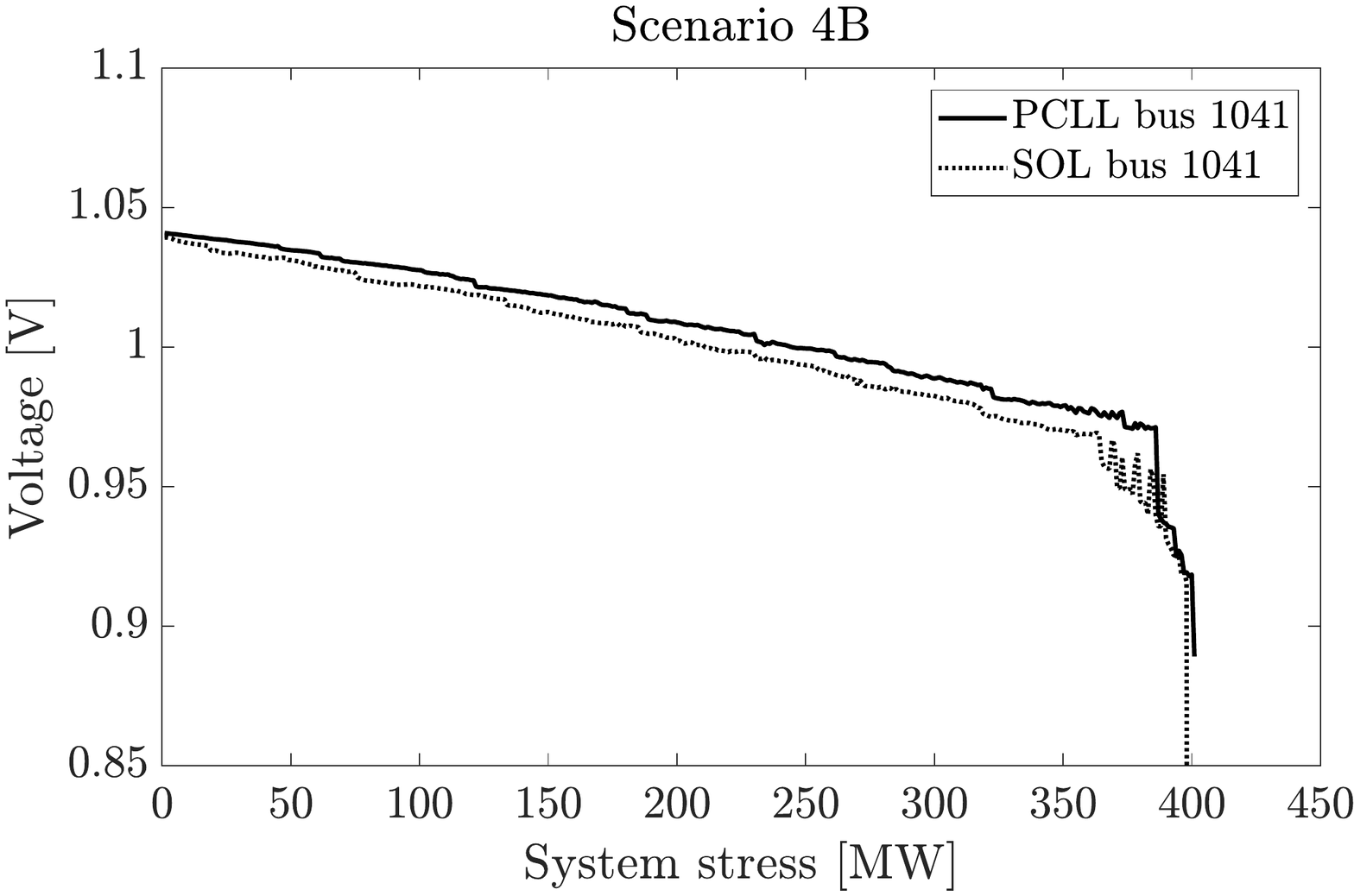}    
		\vspace{-3.1cm}
		\caption{$P\textrm{-}V$ curves computed at bus 1041 for scenario 4B}
		\label{fig:1041_Scenario4B}
		\vspace{-0.85cm}
	\end{center}
\end{figure}

The difference between the PCLL and the SOL is more significant for Scenario B when a longer fault clearing time was used in the simulations. For instance, in Scenario 1B, the SOL was estimated to only 5 MW, compared to 373 MW for the PCLL. With reference to the discussion with the transient $P\textrm{-}V$ curves presented in Section \ref{sec:System_dynamics}, a longer fault clearing time would have the effect of shifting the post-disturbance $P$-$V$ curve for a longer time to the left, causing the system to lack attraction towards a stable post-disturbance equilibrium. Yet again, the difference between the two security margins decreases rapidly as the share of constant active power loads decrease. For instance, in Scenario 4B with an 80\% share constant active power load, and the remaining part of the active load being of constant current characteristics, the SOL and the PCLL are almost identical. The post-disturbance $P\textrm{-}V$ curves of scenario 4B on the transmission side of bus 1041 are illustrated in Fig.~\ref{fig:1041_Scenario4B}. The figure shows that although the computed $P\textrm{-}V$ curves of the SOL is slightly below that of the PCLL, the two security margins find almost the same critical point of the system. 

For all cases, except when the load is of constant power characteristics, the $P\textrm{-}V$ curves computed using the SOL are slightly below the ones computed using the PCLL. Although the initial response of the excitation systems used in the Nordic32 test system is fast, there is an integrating part of the control system which takes a longer time until the voltage magnitudes of the generators are restored to their pre-disturbance set-point (differing slightly due to the droop in the automatic voltage regulation). In the PCLL case, this voltage restoration is allowed to fully stabilize after the initial disturbance before the system stress is added to the system. This is not the case for the SOL, in which the system is stressed before the disturbance is applied to the system. In turn, this causes LTCs and OELs to act earlier for a lower level of system stress, causing the magnitude of the post-disturbance voltages to be generally lower. 

In Scenario C, the chosen contingency was the disconnection of the generator "\textit{g16b}", located in the "\textit{Central}" area. This disturbance was found to more severely affect the stability of the system, and several of the scenarios resulted in negative and/or suddenly unstable simulations. Furthermore, the computed security margins for this case were also found to be sensitive to changes in the simulation setup, such as the used integration time step. Again, the computed SOL was found to be generally smaller or equal than the PCLL for most load configurations, but more care should be used when comparing the results as the difference could be caused by numerical instability during the dynamic simulations.

In Table~\ref{table:Results_CLOD}, the computed SOLs for scenario~A when using different configurations of the CLOD model are presented. The scenarios are generated by varying the load composition, consisting of large induction motors (LIMs), small induction motors (SIMs), discharge lightning (DL), transformer saturation (TS), constant power loads (MVA), and the remaining load (K\textsubscript{p}) which is of constant impedance characteristics. Unsurprisingly, the computed SOL was the lowest when there was a large share of motor loads in the system. When the loads were modelled with a \textit{too} high share of motor loads, such as scenario 16A, the computed SOL for the base case was negative. There was a relatively large difference between the computed SOL for scenario 20A with a 35\% share of LIM loads and 20\% share small motor loads, and scenario 23A with a 20\% share of large motor loads and 35\% share of SIM loads. LIM loads generally draws a higher reactive current during instances of low system voltages than SIM loads, which may have caused the computed SOL to differ from 90 MW for scenario 20A to 196 MW for scenario 23A. 

\begin{figure}
	\begin{center}\vspace{-2.85cm}
		\includegraphics[width=7.2cm]{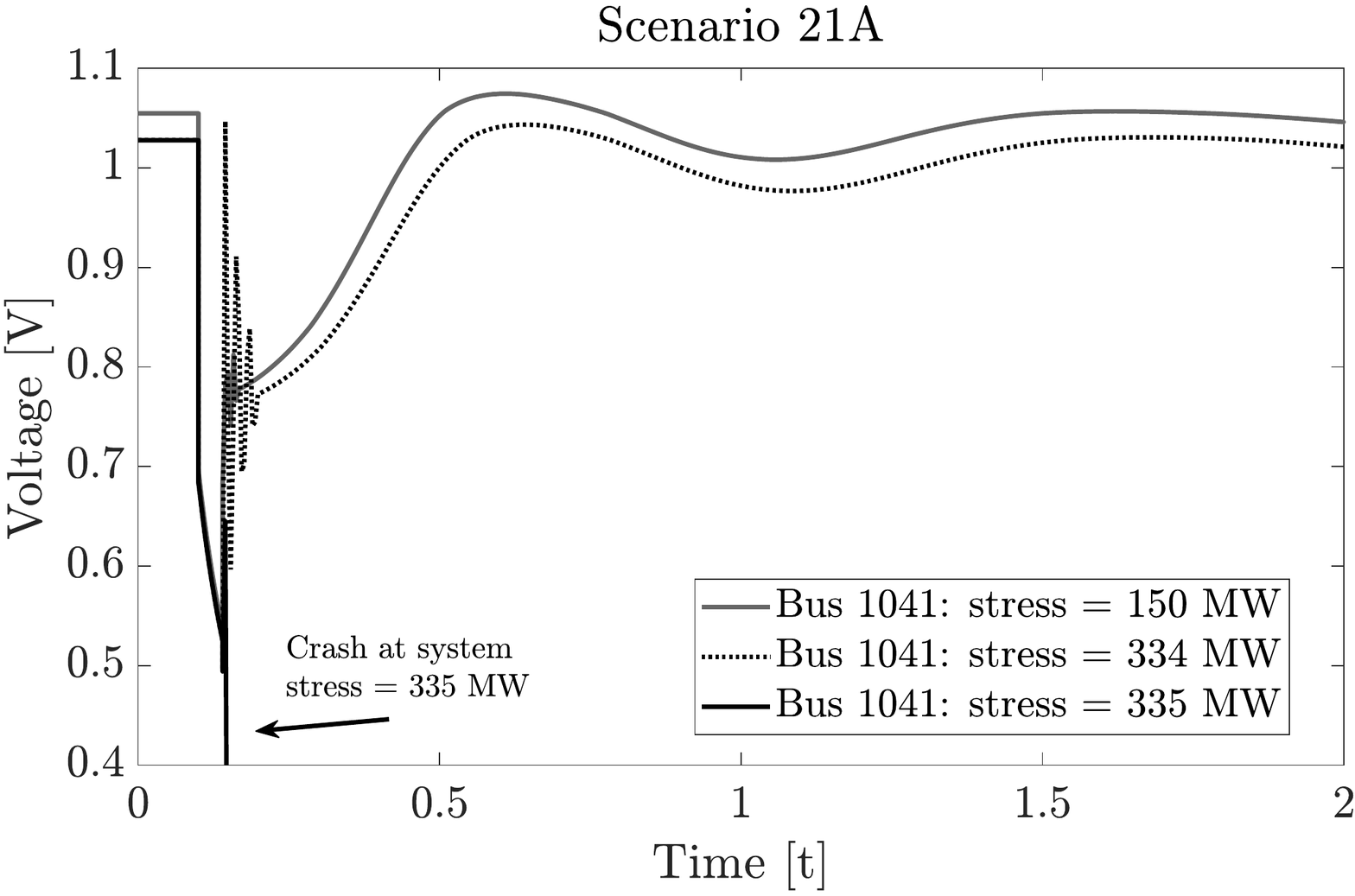}    
		\vspace{-3.1cm}
		\caption{Voltage evolution for bus 1041 for Scenario 23A for different levels of system stress}
		\label{fig:CLOD_Comparison}
		\vspace{-0.8cm}
	\end{center}
\end{figure}

In all scenarios where the CLOD model was used, the system crashed during the transient state just after the disturbance. In Fig \ref{fig:CLOD_Comparison}, scenario 23A is illustrated with varying levels of system stress. For the illustrated lower system stress levels of 150 MW and 334 MW, the system is stable and is capable of handling the dynamics following the disturbance. However, for a stress level of 335 MW, the system crashes almost instantly during the transient state of the disturbance. The CLOD models were found to be particularly sensitive towards long fault clearing times, and the Nordic32 test system consistently crashed when using a longer fault clearing time (such as 0.1 seconds). The difference between the two security margins is thus likely even greater if breakers with longer fault clearing times can be assumed to dominate the system.

\begin{table}
	\vspace{-0.2cm}
	\centering
	\caption{SOLs for different load configurations of the CLOD model}
		\vspace{-0.2cm}
	\label{table:Results_CLOD}
	\begin{tabular}{ccccccccc}
		\toprule
		& \multicolumn{6}{c}{\textbf{Complex load (CLOD) model parameters}} & \textbf{Scenario A} \\ 
		\cmidrule(lr){2-7}\cmidrule(lr){8-8}
        \textbf{Scenario} & LIM & SIM & DL & TS & MVA & Kp & \textbf{SOL} \\ 
	    \textbf{number} & [\%] & [\%] & [\%] & [\%] & [\%] & [MW]\\ 
	    \hline
	    \addlinespace
	    16 & 35 &	35 & 5 & 5 & 10 & 2 &  -\\ 
	    17 & 30	& 30 & 5 & 5 & 10 & 2 &  64\\ 
	    18 & 20	& 20 & 5 & 5 & 10 & 2 &  377\\ 
	    \addlinespace
	    19 & 35	& 30 & 5 & 5 & 10 & 2 &  12\\ 
	    20 & 35	& 20 & 5 & 5 & 10 & 2 &  92\\ 
	    21 & 35 & 10 & 5 & 5 & 10 & 2 &  334\\ 
	    \addlinespace
	    22 & 30	& 35 & 5 & 5 & 10 & 2 & 28\\  
	    23 & 20	& 35 & 5 & 5 & 10 & 2 & 198\\ 
	    24 & 10	& 35 & 5 & 5 & 10 & 2 & 375\\ 
	    \addlinespace
		\bottomrule
		\vspace{-0.6cm}
	\end{tabular}
\end{table}

\subsection{Discussion}
The results in the previous section show that although the \textit{same} operating point has been used as a starting point for all scenarios, the PCLL and the SOL differ significantly depending on the current load configuration and the type of fault that is considered. The largest difference between the two security margin methods was found when either the loads were of high constant power characteristics or consisted of a large penetration of induction motor loads. These results thus confirm the well-known fact that loads with fast restoration dynamics (where a constant power characteristic can be considered a theoretic extreme case) will deteriorate the system stability, and illustrate how significant this impact may be on the computed security margins. However, care should be taken when generalizing the results of this study to real power systems with different characteristics. For instance, although the difference between the SOL and the PCLL in this study was found to be negligible whenever the share of constant power characteristic of the active part of the loads was lower than 50 \%, this is not necessarily the case for other systems with different dynamics. System operators would thus be required to perform a similar analysis on their specific systems to analyze the difference between the PCLL and the SOL. 

The stability assessment practice of many system operators is, to the authors' best knowledge, to compute security margin estimations computed by PCLLs, often in combination with dynamic security assessment (DSA). However, in DSA only the current operating point's capability to withstand a disturbance is assessed, but it provides no information regarding the \textit{margin} towards instability. We believe that if system operators continue to rely on conventional security margins computed by the PCLL, it is important to verify the reliability of those security margins to avoid either overly optimistic security margins or to avoid having to add unnecessary large reliability margins to the computed security margins. To account for modelling inaccuracies, transmission reliability margins are often added to ensure that modelling inaccuracies will not cause the system to be operated unknowingly in a non-secure operating state. Thus, if more accurate methods to determine security margins are used, such as the SOL, these reliability margins may theoretically be reduced and the system operators could more efficiently utilize the existing transmission capacity. 

Dynamic load modelling may also become increasingly important in the future, as more loads are expected to be controlled through power electronically-controlled interfaces. These types of loads, such as electric vehicles chargers, inhibits very fast dynamic responses after disturbances \cite{Mao2018}. Despite this, dynamic load models are still relatively unused in the industry. In a large survey study from 2013 on international industry practice on power system load modelling, it was shown that about 70\% of system operators and utilities around the world still used only static load models for power system stability studies \cite{Milanovic2013}. A drawback of using more advanced load models is that the load composition is often partly unknown to system operators, and it is thus more straightforward to use the simplified static load models. Another drawback is the increase in computational requirement during simulations, which reduces their applicability in real-time applications. However, although complex load models do not necessarily need to be used in real-time applications, sensitivity analyses can preferably be performed using these models, so that the impact of various degrees of motor loads and other types of loads on the stability of a system can be studied. 

\section{Conclusions}
\label{sec:Conclusion}

In this paper, the PCLL and the SOL have been compared and studied under various load configurations and disturbance scenarios. A methodology was developed to allow a fair comparison between the two methods to ensure that the difference in the computed security margins was due to actual differences of the security margin approaches, and not caused by differences in the simulation setups. The numerical comparison shows that the two methods differ significantly under various load configurations and fault scenarios. The largest difference between the two methods was found when the loads were of high constant power characteristics or when the loads consisted of a large share of induction motor loads. Furthermore, the fault clearing time is found to be especially important and a longer fault clearing time caused the SOL to become significantly smaller than the PCLL. The results highlight the importance of load modelling and show that if a power system can be expected to have a large share of loads with fast restoration dynamics, the conventional method of using PCLL to compute the security margins can provide overly optimistic values of the actual security margin.

\bibliography{Referenser}

\begin{thebibliography}{10}
\providecommand{\url}[1]{#1}
\csname url@samestyle\endcsname
\providecommand{\newblock}{\relax}
\providecommand{\bibinfo}[2]{#2}
\providecommand{\BIBentrySTDinterwordspacing}{\spaceskip=0pt\relax}
\providecommand{\BIBentryALTinterwordstretchfactor}{4}
\providecommand{\BIBentryALTinterwordspacing}{\spaceskip=\fontdimen2\font plus
\BIBentryALTinterwordstretchfactor\fontdimen3\font minus
  \fontdimen4\font\relax}
\providecommand{\BIBforeignlanguage}[2]{{%
\expandafter\ifx\csname l@#1\endcsname\relax
\typeout{** WARNING: IEEEtran.bst: No hyphenation pattern has been}%
\typeout{** loaded for the language `#1'. Using the pattern for}%
\typeout{** the default language instead.}%
\else
\language=\csname l@#1\endcsname
\fi
#2}}
\providecommand{\BIBdecl}{\relax}
\BIBdecl

\bibitem{DefinitionVS}
P.~Kundur \emph{et~al.}, ``Definition and classification of power system
  stability ieee/cigre joint task force on stability terms and definitions,''
  \emph{IEEE Trans. Power Syst.}, vol.~19, no.~3, pp. 1387--1401, Aug 2004.

\bibitem{Cutsem1998}
T.~V.~Cutsem and C.~Vournas, \emph{Voltage stability of electric power
  systems}.\hskip 1em plus 0.5em minus 0.4em\relax Boston: Kluwer Academic
  Publishers, 1998.

\bibitem{Cutsem1999}
T.~{Van Cutsem}, C.~{Moisse}, and R.~{Mailhot}, ``Determination of secure
  operating limits with respect to voltage collapse,'' \emph{{IEEE Trans. Power
  Syst.}}, vol.~14, no.~1, pp. 327--335, Feb 1999.

\bibitem{CUTSEM2005b}
T.~{Van Cutsem}, M.-E. Grenier, and D.~Lefebvre, ``Combined detailed and quasi
  steady-state time simulations for large-disturbance analysis,''
  \emph{International Journal of Electrical Power \& Energy Systems}, vol.~28,
  no.~9, pp. 634 -- 642, 2006.

\bibitem{Sittithumwat2002}
A.~{Sittithumwat} and K.~{Tomsovic}, ``Dynamic security margin estimation using
  artificial neural networks,'' in \emph{IEEE Power Engineering Society Summer
  Meeting,}, vol.~3, Chicago, IL, July 2002, pp. 1322--1327.

\bibitem{AMJADY2003}
N.~Amjady, ``Dynamic voltage security assessment by a neural network based
  method,'' \emph{Electric Power Systems Research}, vol.~66, no.~3, pp. 215 --
  226, 2003.

\bibitem{VAKILBAGHMISHEH2007}
M.~V. Baghmisheh and H.~Razmi, ``Dynamic voltage stability assessment of power
  transmission systems using neural networks,'' \emph{Energy Conversion and
  Management}, vol.~49, no.~1, pp. 1 -- 7, 2007.

\bibitem{Hagmar2020IET}
H.~Hagmar, R.~Eriksson, and A.~T. Le, ``Fast dynamic voltage security margin
  estimation: concept and development,'' \emph{IET Smart Grid}, vol.~3, no.~4,
  04 2020.

\bibitem{Dobson1994}
I.~Dobson, ``The irrelevance of load dynamics for the loading margin to voltage
  collapse and its sensitivities,'' in \emph{Bulk power system voltage
  phenomena - III: Voltage stability, security \& control}, 1994, pp. 509--518.

\bibitem{Chowdhury2000}
B.~H. Chowdhury and C.~W. Taylor, ``{Voltage stability analysis: V-Q power flow
  simulation versus dynamic simulation},'' \emph{{IEEE Trans. Power Syst.}},
  vol.~15, no.~4, pp. 1354--1359, Nov 2000.

\bibitem{Ajjarapu1992}
V.~{Ajjarapu} and C.~{Christy}, ``The continuation power flow: a tool for
  steady state voltage stability analysis,'' \emph{{IEEE Trans. Power Syst.}},
  vol.~7, no.~1, pp. 416--423, Feb 1992.

\bibitem{Pal1992}
M.~K. {Pal}, ``Voltage stability conditions considering load characteristics,''
  \emph{{IEEE Trans. Power Syst.}}, vol.~7, no.~1, pp. 243--249, Feb 1992.

\bibitem{Arif2018}
A.~{Arif} \emph{et~al.}, ``Load modeling—a review,'' \emph{IEEE Trans. Smart
  Grid}, vol.~9, no.~6, pp. 5986--5999, 2018.

\bibitem{PAGV2}
\emph{PSS®E 34.2.0 Program Application Guide: Volume II}, Siemens Power
  Technologies International, Schenectady, NY, Apr. 2017.

\bibitem{CutsemTestSystem2020}
T.~{Van Cutsem} \emph{et~al.}, ``Test systems for voltage stability studies,''
  \emph{IEEE Trans. Power Syst.}, vol.~35, no.~5, pp. 4078--4087, 2020.

\bibitem{Mao2018}
D.~{Mao}, K.~{Potty}, and J.~{Wang}, ``The impact of power-electronics-based
  load dynamics on large-disturbance voltage stability,'' in \emph{2018 IEEE
  Power Energy Society General Meeting (PESGM)}, 2018, pp. 1--5.

\bibitem{Milanovic2013}
J.~V. {Milanovic} \emph{et~al.}, ``International industry practice on power
  system load modeling,'' \emph{IEEE Trans. Power Syst.}, vol.~28, no.~3, pp.
  3038--3046, 2013.

\end{thebibliography}
\label{sec:Referenser}

\begin{IEEEbiographynophoto}{Hannes Hagmar}
(S'17) received the M.Sc. degree in electric power engineering from Chalmers University of Technology, Gothenburg, Sweden in 2016. Between 2016 to 2017, he worked at RISE Research Institutes of Sweden with research in electric transmission systems and measurement technology. He is currently pursuing a Ph.D. degree at Chalmers University of Technology. His research interest includes power system dynamics and stability, integration of renewables, and machine learning. 
\end{IEEEbiographynophoto}

\begin{IEEEbiographynophoto}{Robert Eriksson}
(SM’16) received the M.Sc. and Ph.D. degrees in electrical engineering from the KTH Royal Institute of Technology, Stockholm, Sweden, in 2005 and 2011, respectively. He held an associate professor position at the Center for Electric Power and Energy, DTU Technical University of Denmark, from 2013 to 2015. He is currently with the Swedish National Grid (Svenska kraftn{\"a}t), Department of Power Systems. He is also an adjunct professor at the KTH Royal Institute of Technology. His current research interests include power system dynamics and stability, automatic control, and HVDC systems.

\end{IEEEbiographynophoto}

\begin{IEEEbiographynophoto}{Le Anh Tuan}
(S’01–M’09) received a M.Sc. degree in energy economics from the Asian Institute of Technology, Bangkok, Thailand, in 1997, and a Ph.D. degree in power systems from Chalmers University of Technology, Gothenburg, Sweden, in 2004. He is currently a Senior Lecturer with the Division of Electric Power Engineering, Department of Energy and Environment, Chalmers University of Technology. His current research interests include power system operation and planning, power market and deregulation issues, grid integration of renewable energy, and plug-in electric vehicles.
\end{IEEEbiographynophoto}

\end{document}